\documentclass[aps,pre,amsmath,amssymb,twocolumn,floatfix]{revtex4}

\usepackage{graphicx}
\usepackage{multirow}
\usepackage{epsfig}
\newcounter{abc}

\input epsf

\newcommand{\be}{\begin{equation}}
\newcommand{\ee}{\end{equation}}
\newcommand{\bea}{\begin{eqnarray}}
\newcommand{\eea}{\end{eqnarray}}

\newcommand{\cp}{{\cal P}_L}
\newcommand{\tx}{\tilde{x}}

\begin{document}

\title{Fluid phase coexistence and critical behaviour from simulations
in the restricted Gibbs ensemble}

\author{Douglas J. Ashton}
\author{Nigel B. Wilding}
\affiliation{Department of Physics, University of Bath, Bath BA2 7AY, U.K.}
\author{Peter Sollich}
\affiliation{Department of Mathematics, King's College London, Strand, London WC2R 2LS, UK}

\begin{abstract} 

The symmetrical restricted Gibbs ensemble (RGE) is a version of the
Gibbs ensemble in which particles are exchanged between two boxes of
fixed equal volumes. It has recently come to prominence because -- when
combined with specialized algorithms -- it provides for the study of
near-coexistence density fluctuations in highly size-asymmetric binary
mixtures. Hitherto, however, a detailed framework for extracting
accurate estimates of critical point and coexistence curve parameters
from RGE density fluctuations has been lacking. Here we address this
problem by exploiting an exact link between the RGE density fluctuations
and those of the grand canonical ensemble. In the sub-critical region we
propose and test a simple method for obtaining accurate estimates of
coexistence densities. In the critical region we identify an observable
that serves as a finite system size estimator for the critical point
parameters, and present a finite-size scaling theory that allows
extrapolation to the thermodynamic limit. 

\end{abstract} 
\maketitle 
\epsfclipon  

\section{Introduction}

In a recent paper Liu et al \cite{LIU2006} have presented a method for
studying density fluctuations in highly size-asymmetric binary fluids. The
method draws on the geometrical cluster algorithm (GCA) of Liu and Luijten
\cite{Liu2004_0} which facilitates rejection-free Monte Carlo simulations of
highly asymmetric mixtures via large-scale collective updates which move whole
groups of particles (both large and small) in a single step. In its basic
form, the GCA is inherently canonical, i.e.\ it conserves the total number of
particles in the system. To enable density fluctuations -- and thus facilitate
the study of phase behaviour -- it was modified in ref.~\cite{LIU2006} so that
cluster updates transfer particles between two subsystems of fixed equal
volumes, thus realizing complementary density fluctuations within each
subsystem. Such a scenario constitutes a special case of the well known Gibbs
ensemble (GE)~\cite{PANAGIO87}, which has been dubbed the restricted Gibbs
ensemble (RGE) \cite{Mon1992,Bruce97} due to the absence of volume transfers
between the boxes. The lack of volume fluctuations in the RGE (which are
incompatible with cluster updates) means that the two subsystems do not remain
at equal pressure -- a fact which complicates the determination of phase
coexistence points.

In this paper we address the issue of how to determine coexistence curve and
critical point parameters within the RGE. For simplicity we perform the
analysis for a single component fluid, but we expect the results to be
directly applicable to the asymmetrical binary case in which the GCA+RGE
method operates \footnote{This will be the case providing the small particles
are treated grand canonically, in which case for a fixed chemical potential of
the small particles, they simply serve to modify the effective interaction
between the large particles}. Our reference point is the well studied grand
canonical ensemble (GCE) for which powerful techniques exist for determining
critical points and coexistence curve properties \cite{wilding1995a}. In order
to develop methods of comparable utility, we appeal to an exact link between
the density fluctuations in the RGE and those of the GCE. This link is
exploited to {\em predict} the properties of simulations performed in the RGE
simulations on the basis of data accumulated from GCE simulations. A detailed
analysis shows, in particular, how one can locate criticality and the
coexistence binodal within the RGE.

\section{Linking the restricted Gibbs and grand canonical ensembles}
\label{sec:bg}

The symmetrical restricted Gibbs ensemble comprises two subsystems (or
boxes) of fixed equal volumes $V_1=V_2=V=L^d$ (with $d=3$ in the
simulations to be considered below) containing respective particle
numbers  $N_1$ and $N_2$ at some imposed temperature $T$. The boxes can
exchange particles subject to the constraint that the total particle
number 
\be
N_1+N_2=N_0
\label{eq:constraint}
\ee
is fixed. The probability of finding such a restricted Gibbs ensemble system in a given
microstate having particle numbers $N_{1,2}$ and coordinates
$\{q\}_{1,2}$ is given by
\be
P^{RG}(N_{1,2},\{q\}_{1,2}|N_0,V,T)=\frac{\exp(-\beta(E_1+E_2))}{Z^{RG}(N_0,V,T)}\;,
\ee
where $\beta=1/k_BT$ is the inverse temperature, while $E_{1,2}$ is the
configurational energy of the respective boxes. From this it follows that the probability of finding the system with
given $N_1$ (and hence prescribed $N_2$) is
\be
P^{RG}(N_1|N_0,V,T)= \frac{Z(N_1,V,T)Z(N_2,V,T)}{Z^{RG}(N_0,V,T)}\:,
\label{eq:gibbsp}
\ee
where $Z(N,V,T)$ is the canonical partition function. 

Now to forge the link to the grand canonical ensemble, we note that
\be
P(N|\mu,V,T)=\frac{Z(N,V,T)\exp(\mu N/kT)}{Z(\mu,V,T)}\:,
\label{eq:gcedef}
\ee
where $P(N|\mu,V,T)$ is the grand canonical probability distribution at
chemical potential $\mu$. Inserting into eq.~(\ref{eq:gibbsp}) yields
\be
P^{RG}(N_1|N_0,V,T)=w P(N_1|\mu,V,T)P(N_2|\mu,V,T)
\label{eq:prod}
\ee
where the transformation is defined for all $N_1\leq N_0$ and the normalization constant, $w$, is given by
\be
w=\frac{\left[Z(\mu,V,T)\right]^2\exp(-\beta\mu N_0)}{Z^{RG}(N_0,V,T)} \:.
\ee

Eq.~(\ref{eq:prod}) is formally exact and provides the necessary link between
the grand canonical and symmetrical restricted Gibbs ensembles \footnote{It is not,
however, a {\em mapping} because one cannot infer $P(N_1)$ from knowledge of
$P^{RG}(N_1)$}. An interesting property of $P^{RG}$ is its independence of the
choice of the chemical potential appearing on the right hand side of
Eq.~(\ref{eq:prod})-- a feature that has its origin in the complementarity
of the fluctuations of $P(N_1)$ and $P(N_0-N_1)$ \footnote{Notwithstanding
this formal independence, the choice of the value of $\mu$ utilized is
nevertheless significant in a {\em practical} sense because it should lead to
a GCE distribution having appreciable weight in the range of $N_1$ for which
the weight of $P^{RG}$ is concentrated}. 

Of course, it is more conventional to work with the number density rather than
the particle number, in which case the transformation takes the form
\be
P^{RG}_L(\rho| \rho_0,T)\cong P_L(\rho|T)P_L(2\rho_0-\rho|T)\:,
\label{eq:densprod}
\ee
where $\rho=N_1/V$ is the single box density, $P_L(\rho|T)$ is its GCE
distribution at temperature $T$, $\rho_0=N_0/2V$ is the fixed overall
density in the RGE,  and $\cong$ means to within an unspecified
normalization constant. For clarity we have suppressed reference to the
arbitrary chemical potential $\mu$ and the constant volume $V$, but have
added a subscript $L$ to emphasize that we nevertheless remain concerned
with systems of finite size.

A central feature of $P^{RG}_L(\rho)$ is its symmetry: the
transformation (\ref{eq:densprod}) involves the product of $P_L(\rho)$
and a reflected and shifted (by $2\rho_0$) version of itself. Quite
generally [and irrespective of the symmetry of $P_L(\rho)$], this
implies that $P^{RG}_L(\rho)$ is {\em symmetric} (i.e.\ even) with a mean
value of $\rho_0$. This symmetry (and the $\mu$ independence of
$P^{RG}_L(\rho)$ that it implies) precludes deployment of the standard tools for
locating phase coexistence and criticality that have been developed in
the context of the GCE \cite{wilding1995a}. Instead bespoke analysis techniques are called for.

In this paper we develop such techniques. Perhaps surprisingly, our strategy
does not involve performing actual RGE simulations. Instead we apply the exact
transformation Eq.~(\ref{eq:densprod}) to independently obtained GCE density
distributions for the Lennard-Jones fluid in order to {\em infer} the
properties of the resulting RGE density distribution. Our rationale for so
doing is that it is well understood how one determines coexistence and
critical point properties from GCE density distributions. By starting from
such, we have a known baseline with which to compare the results of our
analysis of the RGE. Accordingly, within this approach, GCE distributions
serve as input to Eq.~(\ref{eq:densprod}) and $\rho_0$ simply becomes a
parameter of the transformation. As it is varied, for some prescribed $T$, a
spectrum of symmetrical distributions $P^{RG}_L(\rho|\rho_0)$ is generated,
each centred on the respective value of $\rho_0$. In section~\ref{sec:subcrit}
we consider the nature of this spectrum for state points in the neighbourhood
of liquid-vapor coexistence and present a method for extracting coexistence
densities. Thereafter, in section~\ref{sec:critical}, we consider state points
in the vicinity of the liquid-vapor critical point, and perform a finite-size
scaling analysis of $P^{RG}_L(\rho)$ which facilitates accurate estimates of
critical point parameters. To begin with, however, we shall briefly introduce
our model system.

\section{Fluid model}
\label{sec:LJ}

The GCE density distributions for use as input to the transformation
(\ref{eq:densprod}) were obtained via Monte Carlo simulations
\cite{frenkelsmit2002} of a 3d fluid interacting via a Lennard-Jones
(LJ) potential:
\be
\phi(r)=4\epsilon[(\sigma/r)^{12}-(\sigma/r)^6].
\label{eq:LJdef}
\ee 
Here $r$ is the particle separation, $\sigma$ sets the length scale and
$\epsilon$ is the well depth. The potential was truncated at $r_c=2.5\sigma$
and no correction applied, thus allowing comparison with the results of a
previous study performed under the same conditions \cite{wilding1995a}. As is
customary, we shall refer to the reduced temperature $T\equiv 1/(\beta\epsilon)$
rather than the well depth in the simulation results to be described below.
Furthermore, we shall normally express temperature as a multiple of the critical value
$T_c=1.1878(2)$ (as determined in Sec.~\ref{sec:critical}). Further details of
the simulation methodology can be found in ref.~\cite{wilding1995a}.

For the studies of the subcritical region described in Sec.~\ref{sec:subcrit},
a single periodic system of size $L=10\sigma$ was employed. The finite-size
scaling investigation of the critical region, described in
Sec.~\ref{sec:critical} used seven periodic systems of linear extent
$L=10\sigma,\ 12.5\sigma,\ 15\sigma,\ 17.5\sigma,\ 20\sigma,\ 22.5\sigma$,
$25\sigma$. We set $\sigma$ as the unit of length below. 

\section{Subcritical coexistence region: the intersection method}

\label{sec:subcrit}

\begin{figure}
\includegraphics[width=\columnwidth]{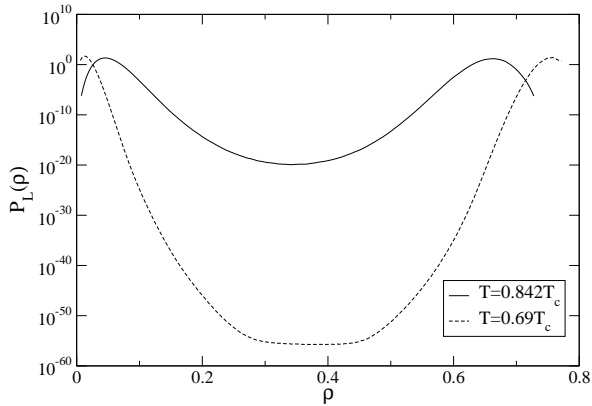}
\caption{The form of the GCE density distribution $P_L(\rho)$ for
temperatures $T=0.842T_c$ and $T=0.69T_c$, expressed on a logarithmic scale.}
\label{fig:GCEdists}
\end{figure}

We commence by addressing the behaviour of the RGE density distributions in the
subcritical coexistence region of our model fluid. As described above, our
approach will be to glean information about the RGE not via explicit RGE
simulations, but by applying the exact transformation Eq.~(\ref{eq:densprod})
to GCE simulation data. Suitable estimates of the GCE density distribution
$P_L(\rho)$ have been obtained in a previous study of the Lennard-Jones fluid by
one of the authors \cite{wilding1995a}. Examples for temperatures $T=0.842T_c$
and $T=0.69T_c$ are shown in Fig.~\ref{fig:GCEdists}, in which the chemical
potential has been tuned so that the peaks have equal areas -- the criterion
for coexistence \cite{Borgs1992}. 

The two peaks in the coexistence form of $P_L(\rho)$ derive from fluctuations
around the respective densities $\rho_g$ and $\rho_l$ of the gas and liquid
phases. These densities are found as an average over the appropriate peak in
$P_L(\rho)$ via
\be \
\rho_{g,l}(T) = \int_{\rho_{\mathrm{min}}}^{\rho_{\mathrm{max}}} \rho  P_L(\rho|T)
\mathrm{d} \rho \:,
\label{eq:peakdens} 
\ee 
where $\rho_{\mathrm{min}}$ and $\rho_{\mathrm{max}}$ are chosen suitably to
fully encompass the appropriate peak of $P^{RG}_L(\rho)$, ie
$\rho_{\mathrm{min}}=\rho_{g,l}-\delta$,
$\rho_{\mathrm{max}}=\rho_{g,l}+\delta$. For our purposes, a useful quantity will prove to be 
the average of the coexistence densities at a given temperature, known as the {\em
coexistence diameter~}: 
\be
\rho_d(T)=[\rho_g(T)+\rho_l(T)]/2\:.
\label{eq:diam}
\ee

A pertinent feature of Fig.~\ref{fig:GCEdists} is the shape of the
distribution in the central region of density, between the peaks. Here one
sees that the distribution flattens markedly, particularly so for the lower
temperature. This flattening corresponds to the appearance of well defined
interfaces between the coexisting phases. For sufficiently low $T$ or
sufficiently large system size $L$, the central portion becomes completely flat because
the interfaces decouple from one another, there being no free energy penalty
for one phase to change its fractional volume with respect to the other
\cite{Errington2003}.

\begin{figure}
\includegraphics[width=\columnwidth]{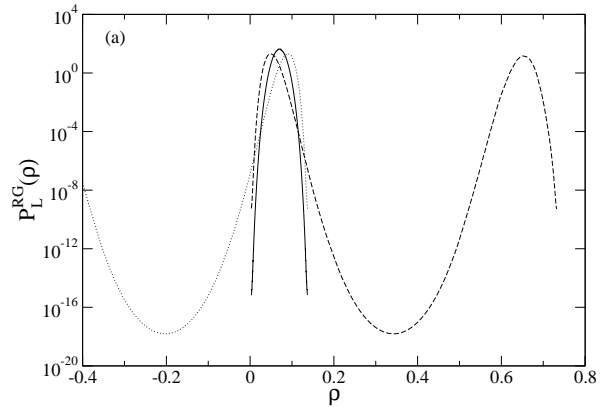}
\includegraphics[width=\columnwidth]{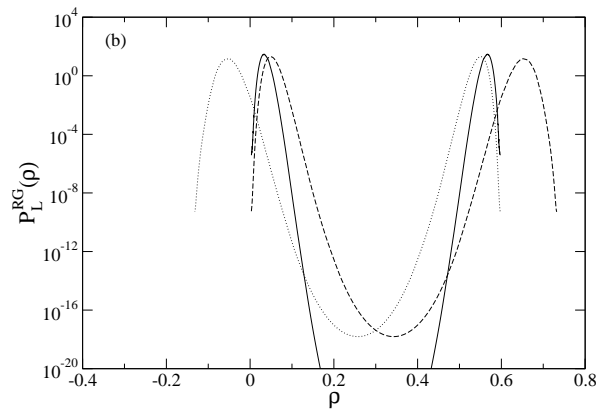}
\includegraphics[width=\columnwidth]{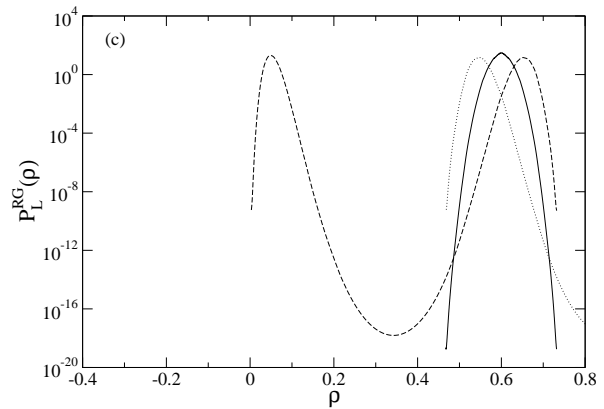}

\caption{The form of $P^{RG}_L(\rho|\rho_0)$ (expressed on a log scale) for the LJ fluid
at $T=0.842T_c$ for {\bf (a)} $\rho_0=0.07$; {\bf (b)} $\rho_0=0.30$;
and {\bf (c)} $\rho_0=0.6$. In each instance the dashed curve is the
coexistence form of the GCE distribution, $P_L(\rho)$; the
dotted curve is the distribution, $P_L(2 \rho_0 - \rho)$, and the solid curve is
the resulting RGE distribution, $P^{RG}_L(\rho)$ obtained via Eq.~(\ref{eq:densprod}).}
\label{fig:rgedists}
\end{figure}

Let us now consider typical forms of $P^{RG}(\rho|\rho_0)$ and how they derive
from the underlying GCE density distribution. Examples are shown in
Fig.~\ref{fig:rgedists} for widely separated values of $\rho_0$. Specifically,
the figure plots the form of $P^{RG}_L(\rho)$ for $T=0.842 T_c$ at three
values of $\rho_0=0.07, 0.30, 0.6$. These show that for the smallest and
highest values of $\rho_0$, $P^{RG}_L(\rho)$ is single peaked, while for the
intermediate value of $\rho_0$, $P^{RG}_L(\rho)$ is double peaked, though in
this latter case neither peak coincides with either the gas or liquid
densities. 

\begin{figure}
\includegraphics[width=\columnwidth]{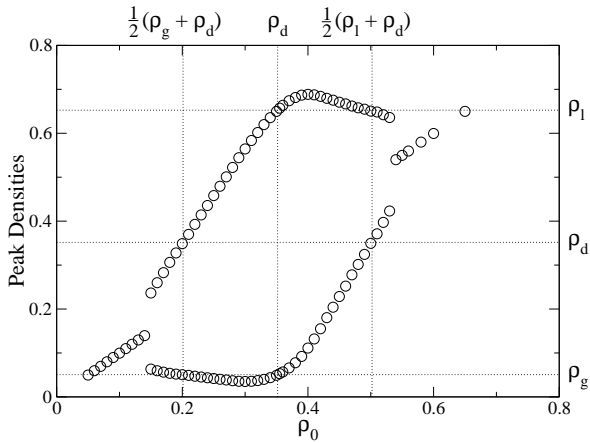}

\caption{The peak densities of $P^{RG}_L(\rho)$ as a function of $\rho_0$ for
the Lennard Jones fluid with $T=0.842T_c$. The dotted lines on the ordinate at
low, intermediate and high densities correspond to the values of $\rho_g$,
$\rho_d$ and $\rho_l$ respectively, as determined independently from the
underlying GCE data as described in the text. The value of $\rho_d$ is also
marked by a vertical dot-dashed line on the abscissa.}

\label{fig:rgepeaks}
\end{figure}

A fuller picture of the $\rho_0$ dependence of $P^{RG}_L(\rho)$ emerges by
considering the peak densities as a function of $\rho_0$. Let us denote the
densities of the low and high density peaks of $P^{RG}_L$ in the double-peak
regime as $\rho_-$ and $\rho_+$ respectively, these being obtained as the
average RGE peak density in a manner analogous to Eq.~(\ref{eq:peakdens}). A plot of
$\rho_-$ and $\rho_+$ versus $\rho_0$ is shown in Fig.~\ref{fig:rgepeaks} for
the temperature $T=0.842T_c$. From this figure one sees that as a function of
$\rho_0$, the RGE distribution is single peaked at low density, splits into
two peaks at intermediate $\rho_0$, but that these peaks merge back into a
single peak at high values of $\rho_0$. Furthermore in the two-peak regime,
$\rho_-$ and $\rho_+$ vary considerably, though we note that they
remain subject to the constraint
\be
\rho_0=(\rho_-+\rho_+)/2\:,
\label{eq:average}
\ee
which is mandated by the symmetry of $P^{RG}_L(\rho)$, together with the fact that 
its average is strictly equal to $\rho_0$.

\begin{table}[h]
\begin{tabular}{cc||c|c|c}

\multicolumn{2}{c}{} & \multicolumn{3}{c}{ {\large $2\rho_0-\rho$}  } \\
\multicolumn{1}{c}{} & &  $\rho_g$ & $\rho_d$ & $\rho_l$ \\
\cline{2-5} \\[-3.1mm] \cline{2-5}

\multirow{3}{*}{{\large $\rho$}~}  &         $\rho_g$& $\rho_g$ & $\frac{\rho_g+\rho_d}{2}$ &$\rho_d$ \\\cline{2-5}
                       & $\rho_d$&         $\frac{\rho_g+\rho_d}{2}$ & $\rho_d$  & $\frac{\rho_l+\rho_d}{2}$\\ \cline{2-5}
                       & $\rho_l$    &     $\rho_d$  & $\frac{\rho_d+\rho_l}{2}$ &  $\rho_l$\\
                      
\end{tabular}
\label{tab:denscomb}
\caption{The combinations of $\rho$ (rows) and $2\rho_0-\rho$ (columns)
corresponding to $\rho_g,\rho_d$ and $\rho_l$. The entries in the table give
the resulting $\rho_0$.}
\end{table}

Now, it transpires that there are certain special values of $\rho_0$ for which
$\rho_-$ and $\rho_+$ both coincide with one of $\rho_g$, $\rho_d$ and
$\rho_l$. To substantiate this, consider the derivative of $P^{RG}(\rho)$.
From Eq.~(\ref{eq:densprod}). This derivative certainly vanishes when
$P_L^\prime(\rho)=P_L^\prime(2\rho_0-\rho)=0$. Assuming $P_L(\rho)$ to be
symmetric, this happens when both $\rho$ and $2\rho_0-\rho$ are one of
$\rho_g$, $\rho_d$, $\rho_l$. The various possibilities are shown in
table~\ref{tab:denscomb} where rows are labelled by $\rho$ and columns
are labelled by $2\rho_0-\rho$, and the entries are the resulting values of
$\rho_0$. Restricting attention to the double peaked regime of $P^{RG}$, one finds from
the table the following three relationships between the peak positions in the RGE and GCE ensembles:

\be
\left.
\begin{array}{ll}
\rho_- &= \rho_g\\
\rho_+ &= \rho_l
\end{array}
\right \} \hspace*{3mm} {\rm when}\:\:\rho_0=\rho_d\:,
\label{eq:atdiameter}
\ee

\be
\left.
\begin{array}{ll}
\rho_- &= \rho_g\\
\rho_+ &= \rho_d
\end{array}
\right \} \hspace*{3mm} {\rm when}\:\:\rho_0=(\rho_g+\rho_d)/2\:,
\label{eq:atlowerdens}
\ee

and 
\be
\left.
\begin{array}{ll}
\rho_- &= \rho_d\\
\rho_+ &= \rho_l
\end{array}
\right \} \hspace*{3mm} {\rm when}\:\:\rho_0=(\rho_l+\rho_d)/2\:.
\label{eq:athigherdens}
\ee

\begin{figure}[h]
\includegraphics[width=\columnwidth]{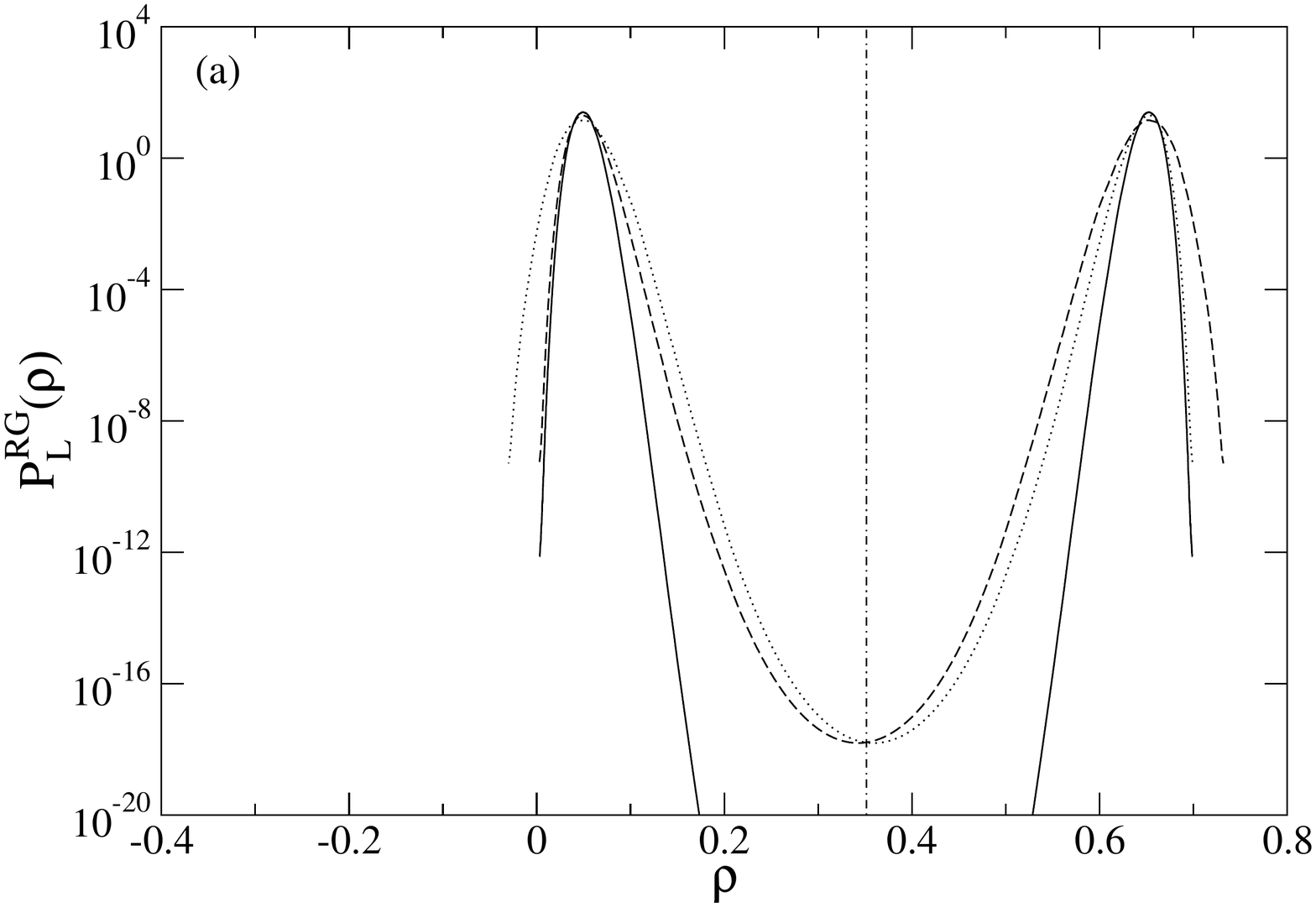}
\includegraphics[width=\columnwidth]{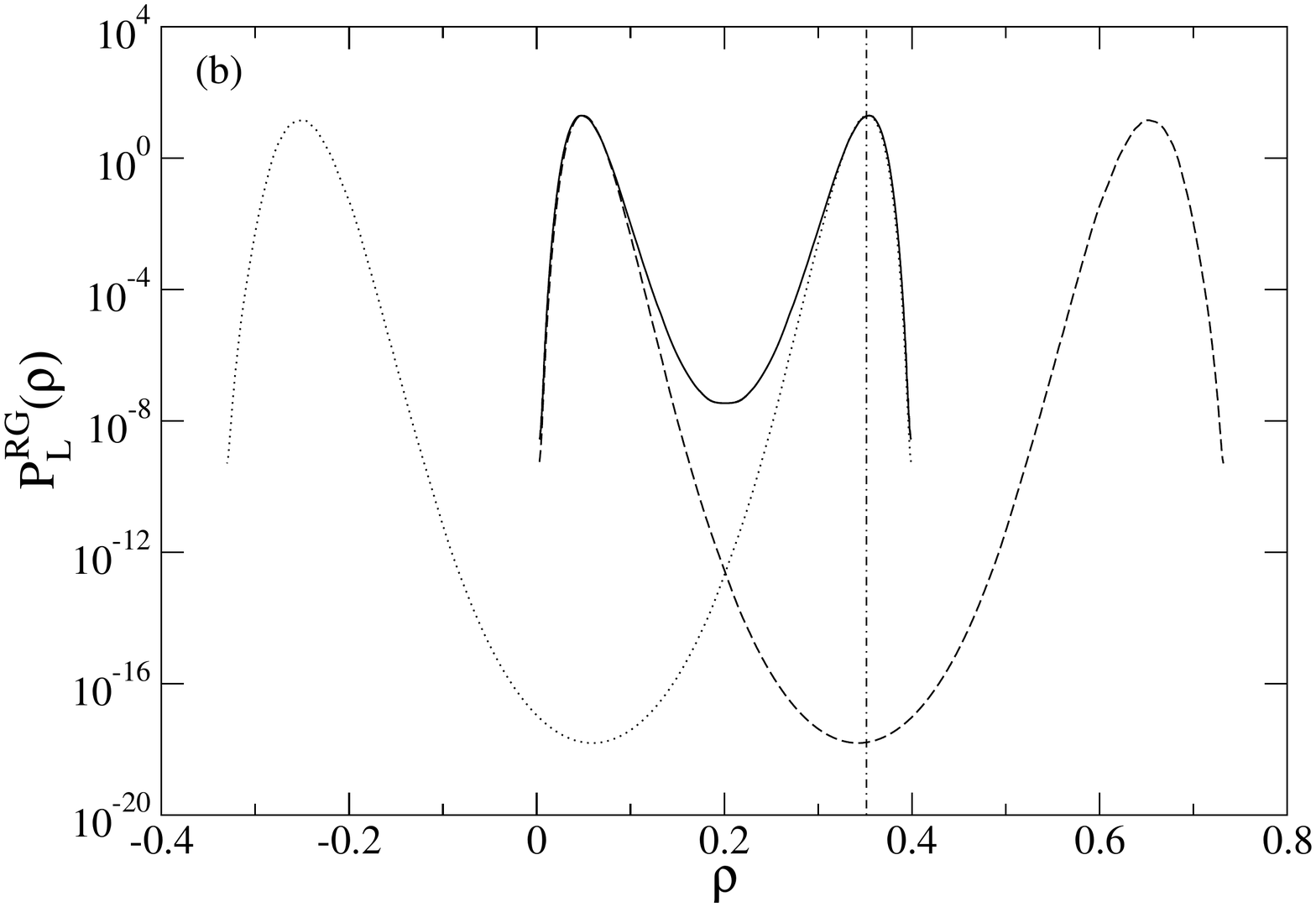}

\caption{{\bf (a).} The solid curve shows the form of $P^{RG}_L(\rho)$
for the case $\rho_0=\rho_d$, for which $\rho_-=\rho_g$ and
$\rho_+=\rho_l$. Also shown (dashed curve) is the underlying GCE
coexistence input distribution $P_L(\rho)$ and the reflected and
shifted distribution $P_L(2\rho_0-\rho)$ (dotted line) that feature in
the transformation Eq.~(\ref{eq:densprod}). {\bf (b).} The corresponding
data for the case $\rho_0=(\rho_g+\rho_d)/2$, for which $\rho_-=\rho_g$
and $\rho_+=\rho_d$. In both parts the coexistence diameter is
represented by a vertical dot-dashed line.}
\label{fig:rgedeconstruct}
\end{figure}

To demonstrate that these predictions are consistent with the data of
Fig.~\ref{fig:rgepeaks}, we have marked on the ordinate the coexistence
densities $\rho_g$, $\rho_l$ and the coexistence diameter $\rho_d$, while the
abscissa is marked with the densities
$\rho_0=(\rho_g+\rho_d)/2$,$\rho_0=\rho_d$, and $\rho_0=(\rho_l+\rho_d)/2$.
Inspection of the figure confirms the above results. Additional insight and
further confirmation is furnished by comparing the forms of $P^{RG}(\rho)$
with the underlying GCE distributions for $\rho_0=\rho_d$ and for
$\rho_0=(\rho_g+\rho_d)/2$, as shown in Fig.~\ref{fig:rgedeconstruct}.

\begin{figure}[h]
\includegraphics[width=\columnwidth]{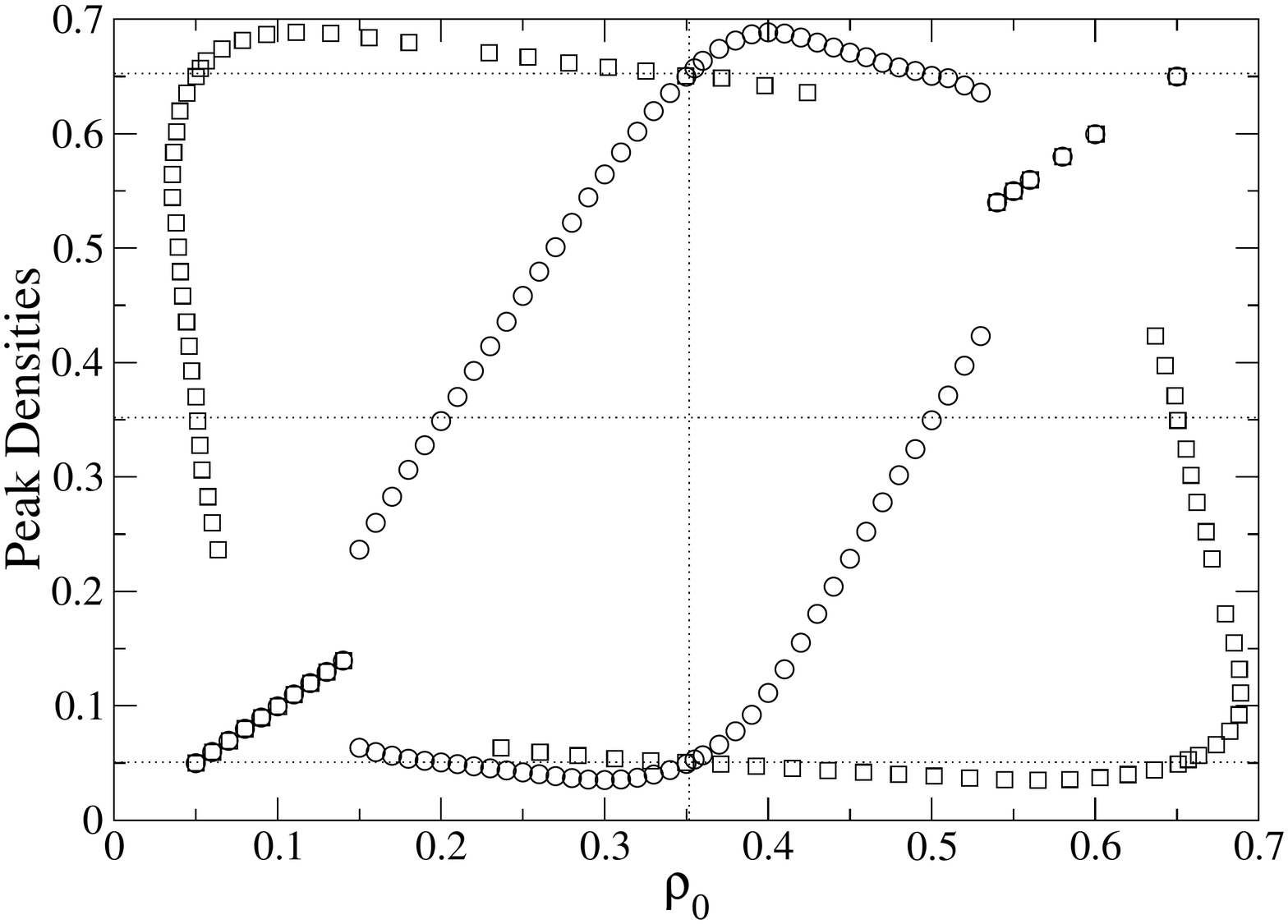}
 \caption{Illustration of the intersection method. Circles show the RGE peak densities
$\rho_-$ and $\rho_+$ of Fig.~\protect\ref{fig:rgepeaks} as a function of $\rho_0$.
Also shown are the transformed points, $2\rho_0-\rho_\pm$, (squares). The
two data sets intersect at the diameter (vertical dot-dashed line).}
\label{fig:intpeaksT}
\end{figure}

Eq.~(\ref{eq:atdiameter}) implies that if one can determine the coexistence
diameter from a RGE simulation, then one can simply read off the coexistence
densities from the corresponding peaks in $P^{RG}_L(\rho|\rho_d)$. In
practice, one can estimate $\rho_d$ by requiring consistency between
Eq.~(\ref{eq:atdiameter}) and Eq.~(\ref{eq:atlowerdens}) and/or
Eq.~(\ref{eq:athigherdens}). The consistency condition is readily implemented
graphically as shown in Fig.~\ref{fig:intpeaksT}: one simply plots the
measured values of $\rho_\pm$ versus $\rho_0$, together with the transformed
points $\rho_\pm$ versus $2\rho_0-\rho_\pm$. The two sets of data intersect at $\rho_0=\rho_d$
and the coexistence densities can be read off from the position of the RGE
peaks at this value of $\rho_0$ \footnote{As
Fig.~(\protect\ref{fig:intpeaksT}) shows, there are two intersections, one for
large peak density and one for small. Close to the critical point, where the
peaks in $P_L(\rho)$ overlap substantially, we observe that the two
intersections occur at slightly different values of $\rho_0$. Empirically,
however, the average of the two intersection densities continues to serve as
an accurate estimate of the coexistence diameter.}. The implementation of this
``intersection method'' is shown in Fig.~\ref{fig:intpeaksT}. 

In order to gauge the accuracy of our intersection method, we tabulate in
Tab.~\ref{tab:diam} the results of applying it at a number of subcritical
temperatures spanning the range $0.7\lesssim T/T_c\lesssim 0.94$. Clearly
there is excellent agreement between the coexistence densities obtained from
the intersection method and those obtained directly from the underlying GCE
coexistence density distribution. Such discrepancies as are evident are
largest at temperatures close to criticality where the positions of peaks in a
density distribution anyway serve as a poor measure of coexistence densities
due to finite-size effects \cite{wilding1995a}.

\begin{table*}[h]
\begin{tabular}{c | c c c | c c c}
 & \multicolumn{3}{c|}{GCE} & \multicolumn{3}{c}{RGE} \\
  $T/T_c$ & $\rho_d$ & $\rho_g$ & $\rho_l$ & $\rho_d$ & $\rho_g$ & $\rho_l$ \\
   \hline
0.937 & 0.3320(16) & 0.1095(12) & 0.554(4) & 0.3352(14) & 0.1096(8) & 0.5608(22) \\
0.865 & 0.3473(24) & 0.0607(8) & 0.6342(44) & 0.3482(20) & 0.0600(10) & 0.6363(44) \\
0.844 & 0.3517(22) & 0.05079(74) & 0.6526(44) & 0.3513(24) & 0.04998(66) & 0.6527(42) \\
0.833 & 0.3538(26) & 0.0467(8) & 0.6611(50) & 0.3537(22) & 0.04598(50) & 0.6615(42) \\
0.813 & 0.3582(12) & 0.0396(12) & 0.6769(26) & 0.3583(16) & 0.0390(4) & 0.6777(32) \\
0.796 & 0.3619(30) & 0.0342(2) & 0.6896(60) & 0.3616(26) & 0.03378(30) & 0.6895(54) \\
0.695 & 0.3847(12) & 0.01360(8) & 0.756(2) & 0.3854(14) & 0.013698(90) & 0.7572(26) \\
\end{tabular}
\label{tab:diam}

\caption{Estimates of the coexistence densities $\rho_g$ and $\rho_l$,
together with the coexistence diameter $\rho_d$ for the 3D LJ fluid described
in Sec.~\ref{sec:LJ} at a selection of sub-critical temperatures. The GCE
estimates of $\rho_{g,l}$ derive from the densities of the peaks in the
coexistence form of $P_L(\rho)$; $\rho_d$ follows as their average. The
RGE estimates for $\rho_d$ derive from $P^{RG}_L(\rho)$ via the intersection method
described in the text, while $\rho_{g,l}$ follows from the corresponding peak
densities in $P^{RG}_L(\rho|\rho_0=\rho_d)$. Error bars are conservative and
are calculated via a blocking analysis of the underlying GCE data.}

\end{table*}

Finally in this section, we note that in the regime of system size and
temperature for which a minimum occurs between the peaks of $P_L(\rho)$
(rather than a flat portion), the rigorous validity of
Eqs.~(\ref{eq:atlowerdens}) and (\ref{eq:athigherdens}) rests upon the
coincidence of the coexistence diameter with the density of this minimum,
i.e.\ on the symmetry of $P_L(\rho)$. In general we observe very close, but
not perfect, agreement between these two densities. However, in practice it
seems that any deviation of the two quantities will impact little on the
effectiveness of the intersection method. This is because near its minimum,
$P_L(\rho)$ is generally much more slowly varying than near its peaks. To
elaborate, consider the case $\rho_0=(\rho_g+\rho_d)/2$ shown in
Fig.~\ref{fig:rgedeconstruct}(b). From the figure one sees that the lower
density RGE peak is formed by multiplying the peak in $P_L(\rho)$ by the
minimum in $P_L(2\rho_0-\rho)$. As the system size increases, the peak becomes
sharper, the minimum becomes flatter, and hence any discrepancy between the
density of the minimum and that of the coexistence diameter will have
increasingly less influence on the location of the low density peak of
$P^{RG}_L(\rho)$. The same argument holds for the high density peak of $P_L^{RG}(\rho)$ at
$\rho_0=(\rho_g+\rho_d)/2$. Of course, in the limit in which the central range
of $P_L(\rho)$ becomes flat, the values of $\rho_-$ (and $\rho_+$) will become
independent of $\rho_0$ for a range of values of $\rho_0\approx\rho_d$. Under
these circumstances the coexistence densities can simply be read off as the
values of these constant densities.


\section{Near-critical finite-size scaling}
\label{sec:critical}

\subsection{Symmetric fluids}

The near-critical finite-size scaling properties of the RGE have been
considered in two previous studies. The first by Mon and Binder \cite{Mon1992}
focused attention on determining the critical temperature for a lattice gas
model, the critical density being known a-priori by virtue of particle-hole
symmetry. Whilst successful with respect to its aims, this study provided no
insight into how to determine the critical parameters for {\em off-lattice}
fluids. The second study, by Bruce \cite{Bruce97}, provided elucidation of the
scaling form of the critical point density distribution (and its relationship
to the critical Ising magnetisation distribution), but here too no method was
given for locating criticality in realistic fluids within the RGE.
 
Here we build on these studies by providing a simple prescription for locating
critical points for off-lattice fluids. To begin with, however, we shall
follow Mon and Binder's lead by seeking inspiration from a model exhibiting
particle-hole symmetry --namely the Ising lattice-gas model -- and consider
the fluctuations in the density $\rho$. We shall first discuss the critical
limit before moving on to the near-critical region.

\subsubsection{The critical limit}

\label{sec:critlim}

In the critical limit $t\equiv (T-T_c)/T_c\to 0$,
$h\equiv(\mu-\mu_c)/\mu_c\to 0$,  and for sufficiently large $L$,
$P_L(\rho)$ adopts the well known universal scaling form \cite{Binder1981}
\be P_L(\rho|\mu_c,T_c)=a_Lp^\star(a_L(\rho-\rho_c))\:.
\label{eq:Isingscale} 
\ee 
Here $a_L=a_1L^{\beta/\nu}$, with $a_1$ a non-universal scale factor,
while $\beta\approx 0.326$ and $\nu\approx0.630$ are the usual critical
exponents for the order parameter and the correlation length
respectively \cite{FISHER98}. $p^\star$ is a universal scaling function,
which is symmetric (even) with respect to the mean density $\rho_c$. Its
form is characteristic of the Ising universality class, and is well
established from previous simulations studies of the Ising model in both
$d=2$ and $d=3$ \cite{NICOLAIDES88,TSYPIN00}.

Let us now consider the RGE density distribution $P^{RG}_L(\rho|\rho_0,T_c)$,
and make the {\em choice} $\rho_0=\rho_c$. Then, from the symmetry of the
critical grand canonical density distribution, we have that
$P_L(\rho|\mu_c,T_c)= P_L(2\rho_0-\rho|\mu_c,T_c)$. Utilizing
Eqs.~(\ref{eq:densprod}) and (\ref{eq:Isingscale}) now shows that 
\be
P^{RG}_L(\rho|\rho_c,T_c) \cong \left[p^\star(a_L(\rho-\rho_c))\right]^2, 
\ee
up to an unspecified normalization factor. Thus the critical point RGE
distribution matches the square of the known universal fixed point form of the
Ising magnetisation distribution, as also previously pointed out by Bruce
\cite{Bruce97} .

\subsubsection{Deviations from criticality}

At the critical temperature $T=T_c$, the form of $P^{RG}$ for values of
$\rho_0$ in the vicinity of $\rho_c$ can be related to that for
$\rho_0=\rho_c$ by expanding in powers of a reduced density
$\varrho_0\equiv\rho_0-\rho_c$. To facilitate ready comparison of the various
members of the resulting spectrum, it is further convenient to work with a
version of $P^{RG}(\rho)$ having zero mean (recall that $\rho_0$ not only
controls the form of this distribution, but also sets its mean). To this end
we set $x\equiv\rho-\rho_c$ and define a shifted GCE density
distribution 
\be
{\cal P}_L(x|\mu_c,T_c)=P_L(x+\rho_c|\mu_c,T_c)\:,
\label{eq:shiftden}
\ee 
so that
\be
P^{RG}(x| \rho_0,T_c)  = \cp(x+\varrho_0|\mu_c,T_c)\cp(-x+\varrho_0|\mu_c,T_c)\:.
\label{eq:rhotransshift}
\ee
Expanding in the small parameter $\varrho_0$, one then finds quite generally
that 
\begin{eqnarray}
P^{RG}_L(x| \rho_0,T_c) & = &\cp(x)\cp(-x) \nonumber\\
                         & + & \varrho_0 \left[  \cp(x)\cp^\prime(-x)+ \cp(-x)\cp^\prime(x) \right]\nonumber\\
                         & + & \frac{1}{2}\varrho_0^2\left[\cp(x)\cp^{\prime\prime}(-x)+2\cp^{\prime}(x)\cp^\prime(-x)\right.\nonumber\\
                         & \: & +\left. \cp(-x)\cp^{\prime\prime}(x)\right]\nonumber\\
                         & + & \frac{1}{6}\varrho_0^3\left[3\cp^\prime(x)\cp^{\prime\prime}(-x)+ 3\cp^{\prime}(-x)\cp^{\prime\prime}(x)\right.\nonumber\\
                         & \: & +\left.\cp(x)\cp^{\prime\prime\prime}(-x)+\cp^{\prime\prime\prime}(x)\cp(-x) \right]\nonumber\\
                         & + & \frac{1}{24}\varrho_0^4\left[8\cp^{\prime\prime}(x)\cp^{\prime\prime}(-x)+\dots\right]+\dots\:,\nonumber\\
\label{eq:genexpan}
\end{eqnarray}
where we have abbreviated $\cp(x)=\cp(x|\mu_c,T_c)$. 

For models exhibiting particle-hole symmetry, the terms in the expansion
(\ref{eq:genexpan}) that involve odd powers of $\varrho_0$ drop out
\footnote{This can perhaps best be appreciated if one sets $\mu=\mu_c$,
in which case $\cp(x|\rho_c)$ is clearly even in $x$. However owing to the
$\mu$-independence of the transformed distribution, the finding is more
generally true.} and one has
\begin{eqnarray}
\label{eq:Isingexpand}
P^{RG}_L(x| \rho_0,T_c) & = & \cp(x)\cp(-x) \nonumber\\
                      & + & \varrho_0^2\left[\cp(x)\cp^{\prime\prime}(x)-(\cp^\prime(x))^2
\right]+{\mathcal O}[\varrho_0^4]\:.\nonumber\\
\end{eqnarray}
Thus in the symmetrical case, the leading $\rho_0$ dependence of $P^{RG}_L(x)$
comes from the $\rho_0^2$ term. An analogous expansion of $P^{RG}_L(x)$ in terms of the reduced
temperature $t$ shows that, by contrast, the leading temperature
dependence arises from terms {\em linear} in $t$. 

Together these findings lead us to propose the following finite-size
scaling {\em ansatz} for $P^{RG}_L(x| \rho_0,T)$, expressed in terms
of the reduced variables $\varrho_0$ and $t$:
\be
P^{RG}_L(x|\varrho_0,t)=b_L{\hat p}(b_Lx,b_2tL^{1/\nu},b_3\varrho_0^2L^{2\beta/\nu},b_4L^{-\theta/\nu})\:.
\label{eq:ansatz}
\ee
Here $b_L=b_1L^{\beta/\nu}$, while $b_1,b_2,b_3,b_4$ are non-universal
scale factors. ${\hat p}$ is a universal scaling function which is symmetric in
$x$ for {\em all} values of its other arguments. The final term, with associated
exponent $\theta\approx 0.52$ \cite{GUIDA98} characterizes the leading
symmetrical corrections to scaling and accounts for deviations from the large
$L$ scaling limit.

Now, a useful dimensionless quantity that characterizes the shape of a
distribution is the fourth order cumulant ratio, a common variant of
which is defined as 
\be
Q\equiv\frac{\langle x^2\rangle^2}{\langle x^4\rangle}\:.
\label{eq:qm0}
\ee
The scaling of the cumulant ratio for $P^{RG}_L(x|\varrho_0,t)$ follows
from Eq.~(\ref{eq:ansatz}) as
\be
Q_L(\rho_0,t)=q(c_2tL^{1/\nu},c_3\varrho_0^2L^{2\beta/\nu},c_4L^{-\theta/\nu})\:,
\ee
with $q$ a universal function. Accordingly in the vicinity of the
critical point the cumulant behaves as
\begin{eqnarray}
Q_L(\rho_0, t)&=&q^\star  \left[    1 + c_2^\prime tL^{1/\nu}+c_3^\prime\varrho_0^2L^{2\beta/\nu}\right. \nonumber\\
           &\ & +\left. c_4^\prime L^{-\theta/\nu}+ \mathcal{O}(\varrho_0^4,t^2)  \right] \:,
\label{eq:Qexpansion}
\end{eqnarray}
where $q^\star\equiv q(0,0,0)$ and
$c_2^\prime,c_3^\prime$ and $c_4^\prime$ are system-specific constants.

\begin{figure}
\includegraphics[type=pdf,ext=.pdf,read=.pdf,width=1.0\columnwidth,clip=true]{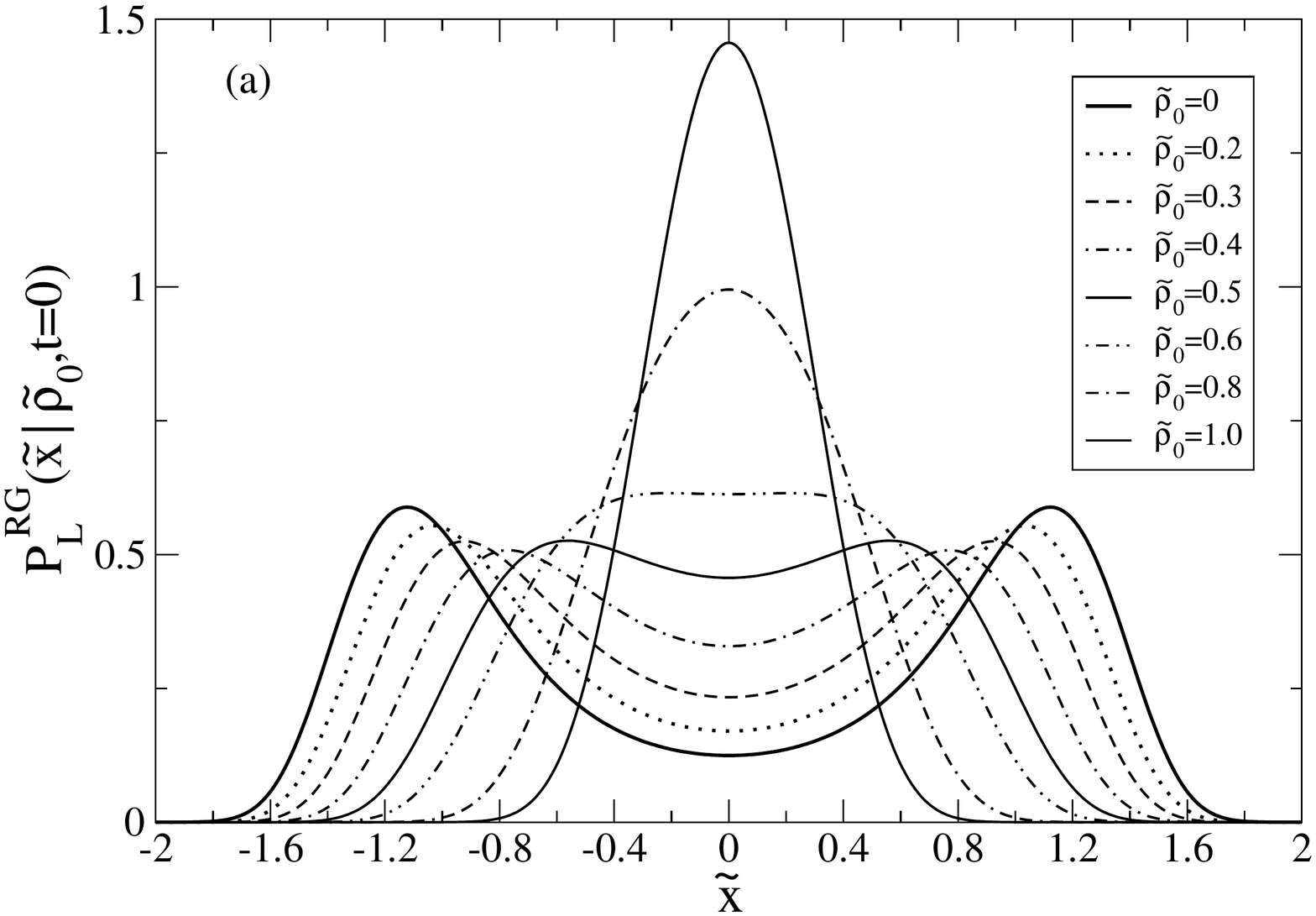}
\includegraphics[type=pdf,ext=.pdf,read=.pdf,width=1.0\columnwidth,clip=true]{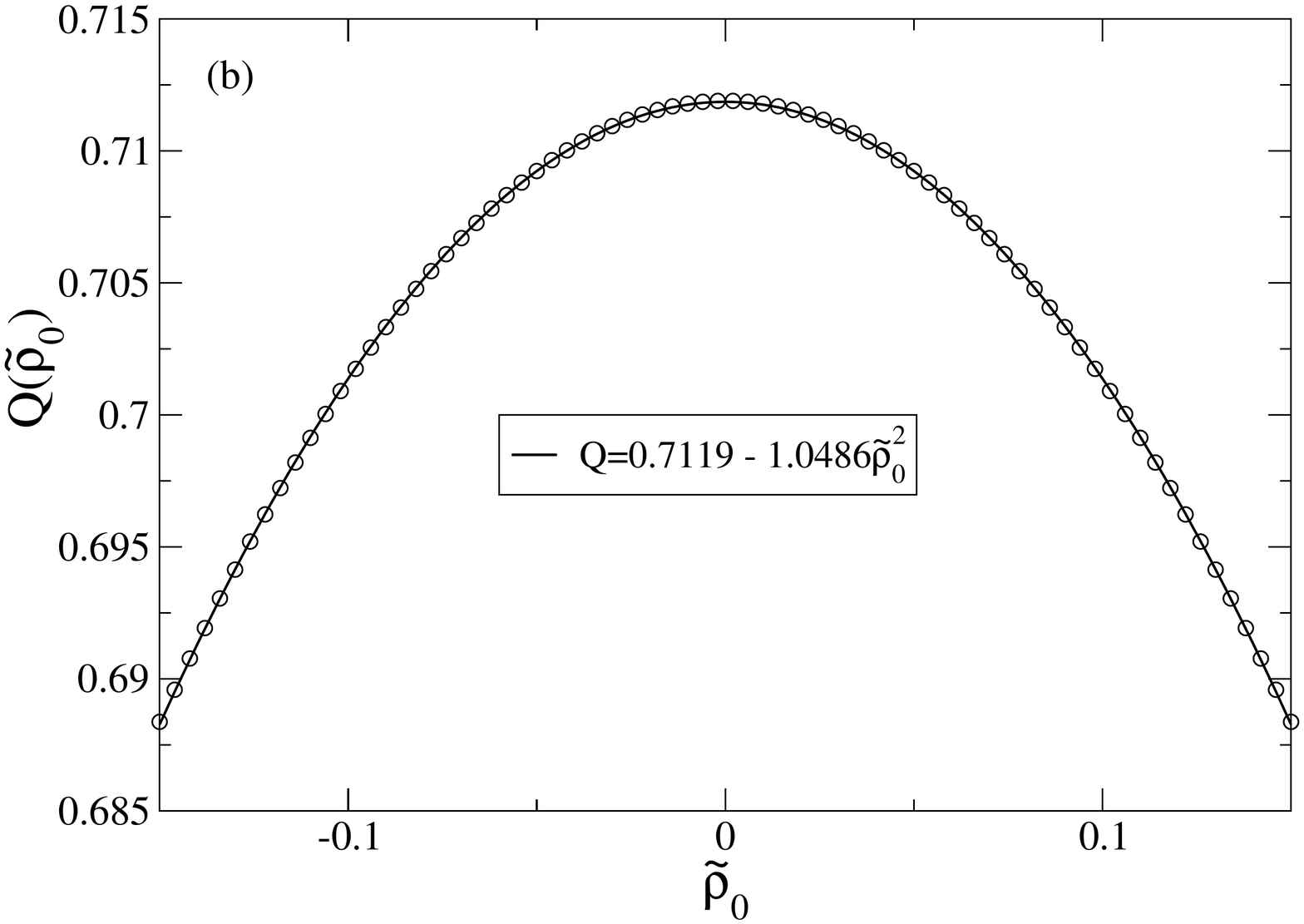}

\caption{{\bf (a)} Selections from the universal spectrum of transformed
distributions generated at $T=T_c$ via
Eqs.~(\ref{eq:Isingscale}-\ref{eq:rhotransshift}) using the estimate for
$p^\star(\tx)$ parameterized in ref.~\protect\cite{TSYPIN00}. All
distributions are normalised. The scaling variable is
$\tilde{x}=b_1L^{\beta/\nu}x$, with the non-universal constant $b_1$ chosen so
that for distribution for $\varrho_0=0$ has unit variance. {\bf (b)} The
corresponding behaviour of the cumulant $Q(\tilde{\varrho_0})$ together with a
parabolic fit. The scaling variable here and in the legend of (a) is
$\tilde{\varrho}_0=\sqrt{c_3}L^{\beta/\nu}\varrho_0$.}

\label{fig:Tsypin} 
\end{figure}

For a given $L$ and a fixed choice of $T$, this implies that the variation of $Q_L$ with $\rho_0$
is parabolic:
\be
Q_L(\rho_0, t=0)=q^\star\left[{\rm const}.+c_3^\prime L^{2\beta/\nu}\varrho_0^2 \right]\:.
\label{eq:Qscanrho0}
\ee
This latter result can be checked directly for the case $T=T_c$ because the
universal critical point form $p^\star(a_L(\rho-\rho_c))$ has been
parameterized on the basis of accurate measurements of the
magnetisation distribution of the critical $3d$ Ising model
\cite{TSYPIN00}. Fig.~\ref{fig:Tsypin}(a) shows a selection from the
spectrum of transformed distributions $P^{RG}_L(x|\varrho_0,t=0)$ for a wide
range of values of $\varrho_0$. For small values of $\varrho_0$, the
behaviour of  $Q_L(\varrho_0)$ is well described by a parabolic form
with $q^\star=0.711901$, as confirmed by the fit shown in
Fig.~\ref{fig:Tsypin}(b).

Finally in this section, it will prove useful to define a special locus
in $(\varrho_0,t)$ space, namely that for which $Q_L$ matches
$q^\star$, which we dub the ``iso-$q^\star$'' line. Let us initially
disregard corrections to scaling (i.e.\ set $c_4^\prime=0$). Then for points on this
line sufficiently close to criticality, it follows immediately
from Eq.~(\ref{eq:Qexpansion}) that
\be
\varrho_0^2=-(c_2^\prime/c_3^\prime)L^{(1-2\beta)/\nu}t \:,
\label{eq:isoqscale}
\ee
which is a parabola in the space of $\rho_0-T$ whose maximum coincides
with the critical point. Now, the presence of
corrections to scaling means that at the critical point ($t=\varrho_0=0$),
$Q_L$ will differ from its limiting value $q^\star$ by an amount
$c_4^\prime L^{-\theta/\nu}$. Effectively, therefore, in following the
iso-$q^\star$ parabola to its maximum for a finite-sized system, we
miss the critical point by a small deviation. Empirically, one finds
that $c_2^\prime$ and $c_3^\prime$ are negative [cf.\ the effect of finite
$\varrho_0$ in Fig.~\ref{fig:Tsypin}(a)]. Since for Ising-like
systems one finds that $c_4^\prime$ is negative, i.e.\ corrections to scaling decrease $Q_L$ below $q^\star$ at
criticality \cite{NICOLAIDES88,wilding1995a}, it falls to negative values
of $t$ to raise $Q_L$ back up to $q^\star$. This implies that the maximum of
the iso-$q^\star$ line occurs at a temperature $T_c(L)$ which differs
from the true critical temperature according to:
\be
T_c-T_c(L)\propto L^{-(\theta+1)/\nu}
\label{eq:apparent_sym}
\ee
Values of $T_c(L)$ obtained from the maximum of the iso-$q^\star$
parabola for a number of system sizes can thus be extrapolated to yield
an estimate of $T_c$.

\subsection{Asymmetric fluids}

In realistic fluids, the critical point density distribution in the GCE
is genuinely symmetric only in the limit $L\to\infty$. For finite-sized
systems at criticality, 
corrections are expected from the fact that the Ising-like scaling
fields are not identical to $t$ and $h$, but are instead given by
linear combinations of these two variables (``field mixing'') and also
the pressure (``pressure mixing''), with in addition quadratic and
higher-order corrections in the same variables.  The magnitude of
these corrections (relative to that of the limiting form $p^\star$)
decays with $L$
\cite{bruce1992,wilding1992,KIMFISHER04}.

Explicitly, one can confirm by a detailed finite-size
analysis~\cite{Sollich_maybe_2010} the form conjectured
by~\cite{KIMFISHER04}, as an expansion in inverse powers of $L$ and up
to quadratic order in $t$:
\begin{eqnarray}
a_L^{-1}&&\!\!\!\!\!\!\!\!
P_L(\rho|\mu_c,T)\ =\ p^* + j L^{-\beta/\nu} s_{\rm pm}
+ a_4 L^{-\theta/\nu} r_{\rm cs}\\
&&{} + l L^{(1-\alpha-\beta)/\nu}s_{\rm fm}
+ a_2 L^{1/\nu} t\,r_t
\nonumber\\
&&{}- k L^{(2-\alpha-\beta)/\nu} t\,s_{t,\rm fm}
+ k^2 L^{2(2-\alpha-\beta)/\nu} t^2\,r_{t^2,\rm fm}
\nonumber
\end{eqnarray}
Here $p^*$, $r_{\rm cs}$, $r_t$, $r_{t^2,\rm fm}$ are even functions,
while $s_{\rm pm}$, $s_{\rm fm}$ and $s_{t,\rm fm}$ are odd; all are
evaluated at the scaled density $a_L(\rho-\rho_c)$. All terms on the
r.h.s.\ except for $p^*$ have non-universal prefactors which we have
written as $j$, $k$, $l$, $a_2$ and $a_4$. The subscripts indicate
where the various  
contributions come from, according to pm: pressure mixing, cs:
corrections to scaling, fm: field mixing, $t$: temperature shift,
$t^2$: second order in temperature shift.

From $P_L(\rho|\mu_c,T)$ as given above one can find
$P_L^{RG}(x|\rho_0,T)$, and then take moments of $x$ to obtain the
cumulant ratio $Q$. This basically inherits the
structure of the terms in $P_L(\rho|\mu_c,T)$, except for the fact
that the asymmetry of $s_{\rm pm}$ and $s_{\rm fm}$ cancels linear
contributions in the mixing coefficients:
\begin{eqnarray}
Q_L(\rho_0,t)&=& q^*[
1+c_2'tL^{1/\nu}+c_3'\varrho_0^2 L^{2\beta/\nu} + c_4'
L^{-\theta/\nu}\ \ \ \ \ {}
\label{QL_with_mixing}
\\
&&{}+\ldots j^2 L^{-2\beta/\nu}
+\ldots jk L^{(2-\alpha-2\beta)/\nu} t
\nonumber\\
&&{}+\ldots k^2 L^{2(2-\alpha-\beta)/\nu} t^2
\nonumber\\
&&{}+\ldots j \varrho_0 L^{-2\beta/\nu}
+\ldots k \varrho_0 L^{(2-\alpha-2\beta)/\nu}]
\nonumber
\end{eqnarray}
The dots ($\ldots$) indicate factors of order unity that are
determined by the various universal functions $r$ and $s$. It is then
easy to see, by setting $\partial Q_L/\partial \varrho_0=0$,
that the extremum of any line of constant $Q_L$ in the $(\varrho_0,t)$
plane occurs at
\be
\varrho_0 = \ldots j L^{-2\beta/\nu} + \ldots k
L^{(2-\alpha-2\beta)/\nu} t
\label{varrho_max}
\ee
rather than at $\varrho_0=0$ as in our previous analysis without field and
pressure mixing. Inserting back into (\ref{QL_with_mixing}) then shows
that the value of $Q_L$ on any iso-$Q_L$ line has, as a function of the reduced
temperature $t$ at the maximum, the same form as the r.h.s.\ of
(\ref{QL_with_mixing}) but with the $\varrho_0$-dependent terms
omitted and the dots now representing different prefactors of order
unity. By setting 
$Q_L=q^*$ one concludes that the maximum of the iso-$q^*$ curve, in particular,
lies at the reduced temperature obeying
\begin{eqnarray}
0&=&c_2'tL^{1/\nu}+c_4' L^{-\theta/\nu}
+\ldots j^2 L^{-2\beta/\nu}
\label{qstar_t}
\\
&&{}
+\ldots jk L^{(2-\alpha-2\beta)/\nu} t
+\ldots k^2 L^{2(2-\alpha-\beta)/\nu} t^2
\nonumber
\end{eqnarray}
The corresponding value of $\varrho_0$ at the maximum follows by inserting the
solution for $t$ into (\ref{varrho_max}).

We have kept track of pressure mixing effects via the coefficient $j$
above, but in non-ionic fluids these are generally weak, i.e.\ $j$ is
very small~\cite{KIMFISHER04}. To see how 
mixing effects enter, let us then only keep $k$ for now. If $k$ is
small, the condition (\ref{qstar_t}) will
give the leading scaling $t\sim L^{-(\theta+1)/\nu}$ as before. The
last term in (\ref{qstar_t}) is then of order $k^2
L^{2(1-\alpha-\beta-\theta)/\nu}$ and becomes comparable to the first
two terms $\sim L^{-\theta/\nu}$ when
\be
L \sim k^{-\nu/(1-\alpha-\beta-\theta/2)}
\ee
The exponent on the r.h.s.\ is around -2.07, so if $k$ is small enough
then this crossover lengthscale above which field mixing becomes
relevant may be too large for us to access. Our numerical results
below suggest that this is the case for the specific system studied
here.  

In Eq.~(\ref{varrho_max}), the pressure mixing term is generally
subleading because of its $L$-dependence, so that 
$\varrho_0 \sim k L^{(2-\alpha-2\beta)/\nu} t$. Even if mixing is too
weak to be observed in the $L$-dependence of $t$, it will therefore be
evident in an $L$-dependence of $\varrho_0$. In the specific case
where $t\sim L^{-(\theta+1)/\nu}$, this dependence would be 
\be
\varrho_0
\sim k L^{(1-\alpha-2\beta-\theta)/\nu}
\label{varrho_scaling}
\ee
with exponent $\approx -0.45$.

\subsection{Computational method}

Following the iso-$q^\star$ parabola to its maximum provides a convenient
route to the effective critical parameters. In practice, this is achieved via
the following computational prescription.

\begin{enumerate}

\item For a given system size $L$, perform a RGE simulation for some
value of $\rho_0$ and $T$ to obtain $P^{RG}_L(\rho|\rho_0,T)$.

\item Deploy histogram reweighting \cite{ferrenberg1989} to locate the
temperature for which $Q_L(T|\rho_0) = q^\star=0.711901$. 

\item Repeat for a range of $\rho_0$ values to yield the
iso-$q^\star$ curve in $\rho_0-T$ space.

\item A parabolic fit to the estimates for the iso-$q^\star$ curve will
pinpoint its maximum or -- should it prove necessary -- guide the
choice of parameters for further simulations nearer to criticality.

\item The coordinates of the maximum of the iso-$q^\star$ parabola
serve as estimates for $\rho_c(L)$ and $T_c(L)$. Results from a range
of system sizes can be extrapolated to the thermodynamic limit
according to Eqns.~(\ref{varrho_scaling})
and~(\ref{eq:apparent_sym}). 

\end{enumerate}

With regard to this procedure, we make the following observations. Firstly,
our approach explicitly assumes that the universality class and the associated
value of $q^\star$ are known a-priori. In most cases of current interest,
fluid criticality appears to be Ising-like, and hence the above procedure
should provide a precise and efficient procedure for estimating $T_c$ and
$\rho_c$. However, in cases where the universality class is in doubt, or one
simply wishes to perform a consistency check on one's measurements, a
procedure can be implemented analogous to that proposed by Kim and Fisher
\cite{KIMFISHER03} for the GCE density distribution. By scanning $\rho_0$ at
some prescribed $T$, one can locate the maximum of the parabola $Q(\rho_0|T)$
(cf.\ Eq.~(\ref{eq:Qscanrho0})). Repeating for a range of $T$ (via histogram
reweighting) allows the determination of the line of $Q$ maxima in $\rho_0-T$
space. Estimates of $Q$ along this line, and for a range of system sizes,
should exhibit ``cumulant crossing'' \cite{Binder1981} at the estimated
critical point, and for a value of $Q=q^\star$ characteristic of the
appropriate universality class. Correspondingly the finite size forms of
$P^{RG}_L(x)$ should all collapse onto a universal scaling function.

\subsection{Results}

In order to investigate the finite-size scaling properties of the RGE, we have
generated near-critical forms of $P^{RG}_L$ from GCE density distributions via
the transformation Eq.~(\ref{eq:densprod}).  

\begin{figure}
\includegraphics[type=pdf,ext=.pdf,read=.pdf,width=1.0\columnwidth,clip=true]{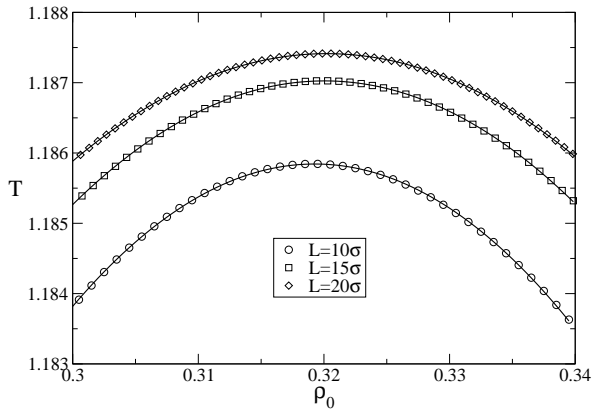}
\caption{Measured iso-$q^\star$ curves for the LJ fluid for system sizes
$L=10,\ 15,\ 20$. Uncertainties are considerably smaller
than the symbol sizes. Also shown (curves) are parabolic fits; maxima provide
estimates of the critical temperature $T_c(L)$ and density $\rho_c(L)$.}
\label{fig:isoQstar}
\end{figure}

Measurements of the iso-$q^\star$ curve for the system sizes $L=10,\
15,\ 20$ are shown in Fig.~\ref{fig:isoQstar}(a) together with a
parabolic fit. Clearly the data are well described by a parabolic form, but
the positions of the maxima vary considerably.

\begin{figure}
\includegraphics[type=pdf,ext=.pdf,read=.pdf,width=1.0\columnwidth,clip=true]{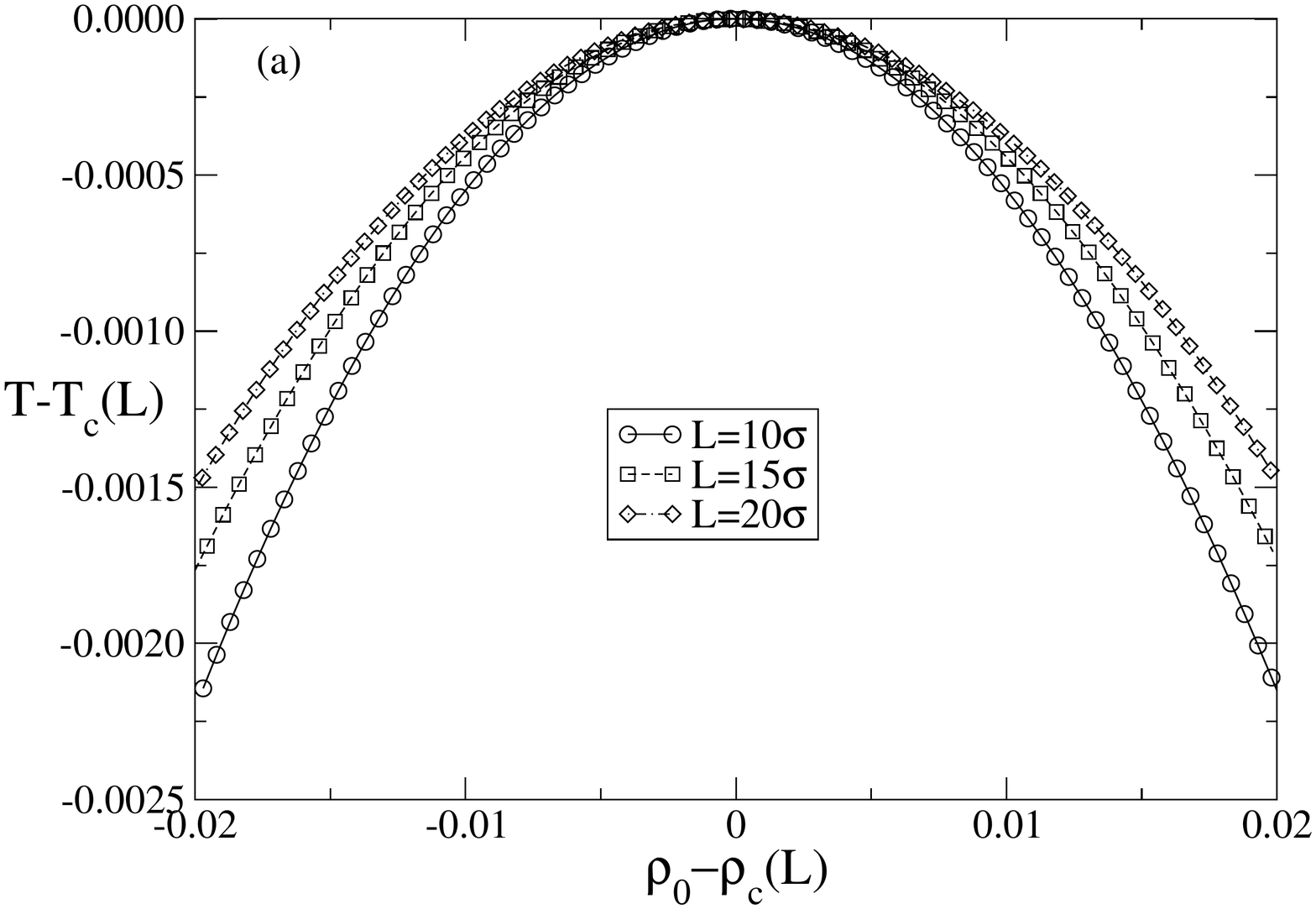}
\includegraphics[type=pdf,ext=.pdf,read=.pdf,width=1.0\columnwidth,clip=true]{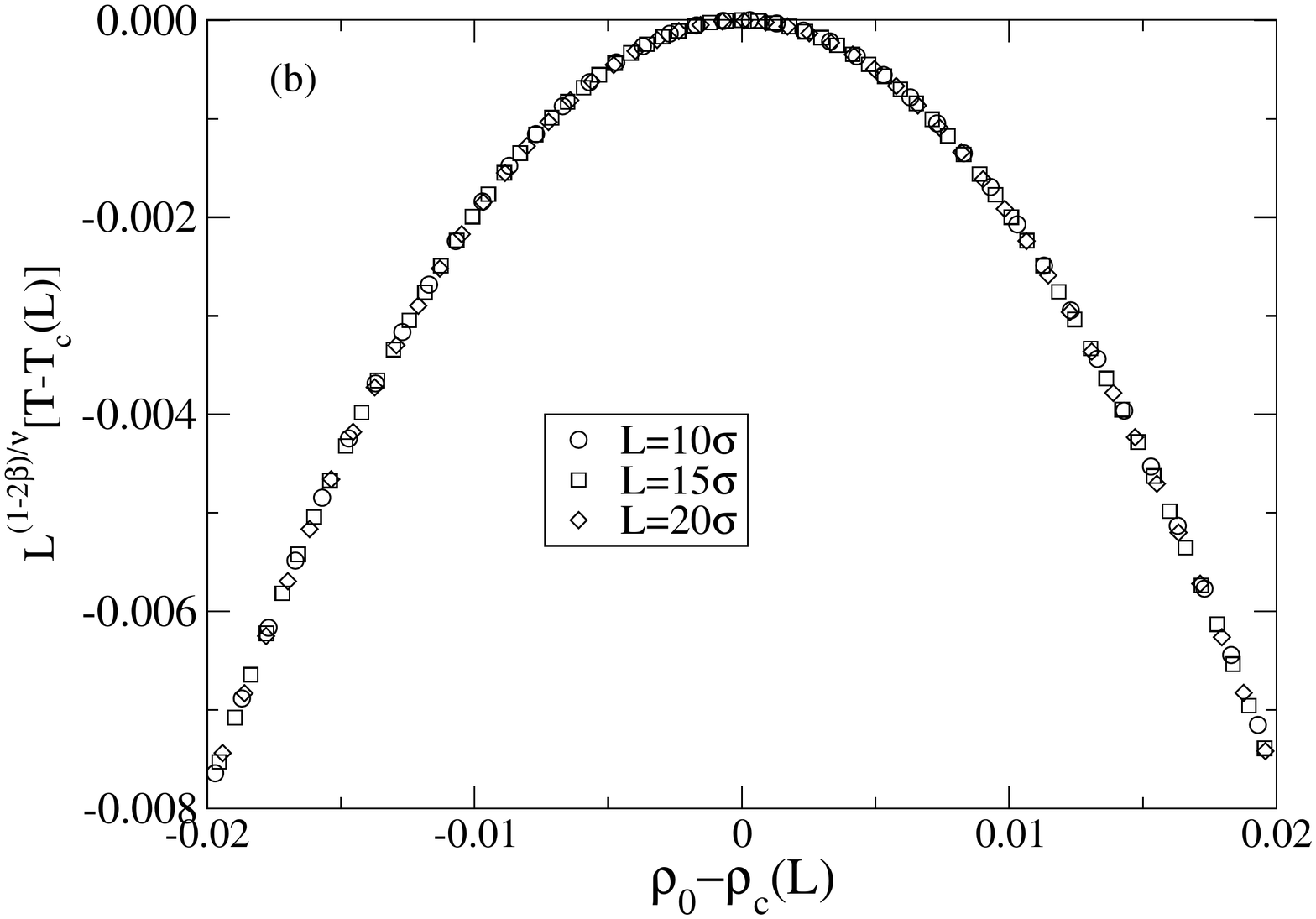}
\caption{{\bf (a)} The iso-$q^\star$ curves of
fig.~\protect\ref{fig:isoQstar} shifted so that their maxima
lie at a common origin. Statistical errors are smaller than the symbol
sizes; lines are guides to the eye. {\bf (b)} The same data plotted in terms of the scaling variable
$L^{(1-2\beta)/\nu}t$, with $\beta=0.326,\nu=0.63$.}
\label{fig:isoQ*scale}
\end{figure}

In order to expose the scaling behaviour of the iso-$q^\star$ curves we
have shifted the data so that the maxima coincide with the origin,
as shown in Fig.~\ref{fig:isoQ*scale}(a). We then scale the temperature axis
by $L^{(1-2\beta)/\nu}t$, as predicted by Eq.~(\ref{eq:isoqscale}) to
obtain the results shown in Fig.~\ref{fig:isoQ*scale}(b). This indeed
shows an excellent data collapse.

\begin{figure}

\includegraphics[type=pdf,ext=.pdf,read=.pdf,width=1.0\columnwidth,clip=true]{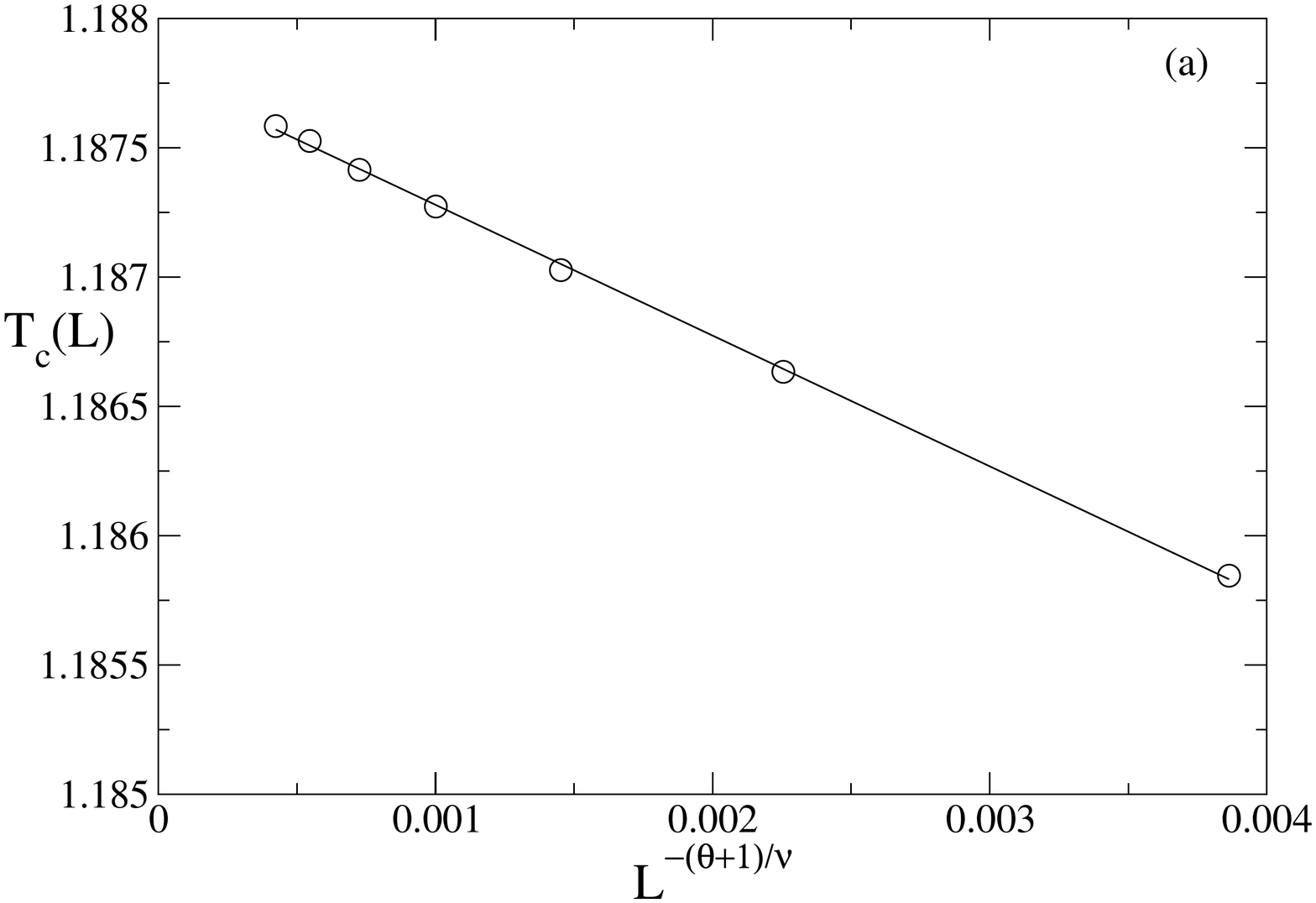}
\includegraphics[type=pdf,ext=.pdf,read=.pdf,width=1.0\columnwidth,clip=true]{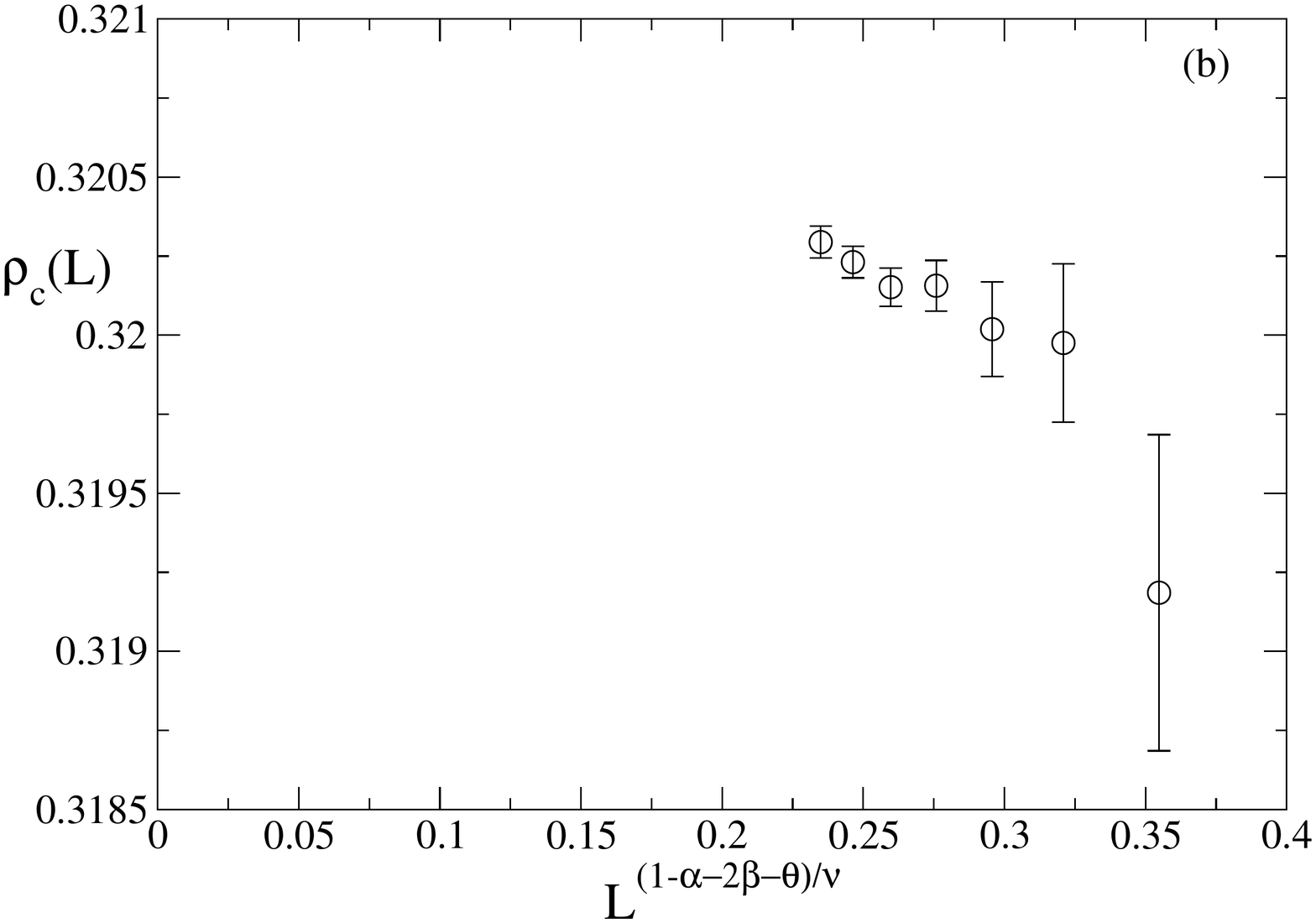}

\caption{{\bf (a)} Measured estimates of $T_c(L)$ from the maxima of the
iso-$q^\star$ curve. The data is plotted against the
scaling variable $L^{(\theta+1)/\nu}$ and a straight line fit shows the
quality of the scaling. Statistical errors do not exceed the symbol sizes.
{\bf (b)} Measured estimates of $\rho_c(L)$ from the maxima of the
iso-$q^\star$ curves, plotted against the anticipated scaling variable,
Eq.~(\ref{varrho_scaling}) }
\label{fig:Tcscaling}
\end{figure}

The coordinates of the iso-$q^\star$ maxima serve as estimates for the
effective critical parameters $T_c(L), \rho_c(L)$. The finite-size dependence
of these estimates are plotted in Figs.~\ref{fig:Tcscaling}(a) and (b)
respectively in terms of the anticipated scaling variables (cf.\
Eqs.~(\ref{eq:apparent_sym}) and~(\ref{varrho_scaling})). As anticipated
above, data for $T_c(L)$ are most consistent with the scaling $t\sim
L^{-(\theta+1)/\nu}$, suggesting that field mixing effects on this quantity
are weak, at least for the system sizes we have studied. An extrapolation of
the data for $T_c(L)$ (Fig.~\ref{fig:Tcscaling}(a)) yields $T_c=1.1878(2)$. As
regards $\rho_c(L)$, the range of accessible system sizes and the rather large
error bars unfortunately prevent us from establishing convincingly the
predicted $L$-dependence (\ref{varrho_scaling}). The error bars were
determined from a blocking analysis and one notes that those for $T_c(L)$ are
relatively much smaller than those on $\rho_c(L)$. The reason for this are
twofold. Firstly the iso-$q^\star$ curves have quite small curvature (cf. the
scales of ~Fig.~\ref{fig:isoQstar}) so that any uncertainty in $T$ translates
to a larger uncertainty in $\rho$. A further problem is that for the smaller
system sizes the resolution of the density scale $\Delta\rho=1/V$ limits the
precision with which the maximum can be determined. For instance for $L=10$,
the density resolution is $\Delta\rho=0.001$ which is large on the scale
of the finite size shifts we observe overall. If we restrict our extrapolation
to a fit based only on the five largest system sizes shown in
Fig.~\ref{fig:Tcscaling}(b), we find $\rho_c=0.3210(3)$. The present estimates
for $T_c$ and $\rho_c$ are to be compared with previous older estimates (based
on a more limited range of system sizes) of $T_c=1.1876(3),\rho_c=0.3197(4)$.
\cite{wilding1995a}.

\section{Summary and conclusions}

The aim of the work described in this paper was to lay firm foundations for
future investigations of phase behaviour within the Restricted Gibbs
ensemble. To this end we have developed new approaches for determining both
the coexistence and critical point parameters within this ensemble. 

In the subcritical regime, we have proposed and tested an ``intersection
method'' for estimating the coexistence densities. This involves measurements
of the RGE peak densities as a function of the overall system density
$\rho_0$. Comparison of the resulting data with a transformed version of itself yields an
intersection at a density that serves as an estimate of the coexistence diameter
$\rho_d$. The coexistence densities are then simply read off from the RGE peak
densities for $\rho_0=\rho_d$. In tests, the results agreed to high precision
with independently determined estimates of the coexistence properties.

In the near-critical region, we have described (and extended) a finite-size
scaling method (elements of which were originally outlined in
ref.~\cite{LIU2006}) for obtaining accurate estimates of fluid critical point
parameters within the RGE. The strength of this method is that it provides
estimates of effective (finite-size) critical parameters via a simple and
convenient parabolic fit to measurements of a cumulant ratio. These estimates
can be extrapolated to the thermodynamic limit using appropriate scalings
forms which we have presented both for the case of particle-hole asymmetry and
for asymmetric fluids. We expect the method to be useful when one has prior
knowledge of the universality class and the associated value of $q^\star$. 

In addition to its utility for extracting critical point parameters from RGE
simulations of fluids, we remark that our analysis provides an alternative to
traditional routes for obtaining these properties in GCE simulations.
Specifically, as we have shown, one can apply the transformation of
Eq.~(\ref{eq:densprod}) to near-critical GCE data and then proceed as if the
density distribution had been obtained within the RGE. The subsequent analysis
in terms of the scaling of iso-$q^\star$ maxima is arguably simpler to implement
than a widely used approach for GCE data \cite{wilding1995a} in which one
determines a symmetrized ordering operator distribution via a linear field
mixing approximation, and matches this distribution to a universal form to
obtain effective critical parameters. 

Finally, with regard to the wider context of this work, it should be stressed
that RGE simulations cannot be regarded as competitive with the GCE in situations
where the latter operates efficiently, e.g.\ for single component fluids or mixtures of
similarly sized particles. Specifically, the lack of a chemical potential field
implies that histogram reweighting cannot be implemented to scan a range of
overall densities or concentrations on the basis of a single simulation run;
instead a separate RGE simulation is required for each density $\rho_0$ of
interest. Where we expect the RGE to come into its own, is for highly
size-asymmetric mixtures, for which GCE sampling becomes ineffectual. As shown
in ref.~\cite{LIU2006} the Generalized Cluster algorithm, when combined with
the RGE, does permit efficient sampling of configuration space in such
systems, and thus the methods we have described should facilitate the accurate
determination of phase coexistence properties. We have already begun applying
them in this context and hope to report on our progress in due course.

\acknowledgments

The authors thank Erik Luijten and Rob Jack for helpful discussions and
suggestions.

\appendix

\bibliographystyle{prsty}

\begin{thebibliography}{10}

\bibitem{LIU2006}
J. Liu, N.B. Wilding, and E. Luijten, Phys. Rev. Lett. {\bf 97},  115705
  (2006).

\bibitem{Liu2004_0}
J. Liu and E. Luijten, Phys. Rev. Lett. {\bf 92},  035504  (2004).

\bibitem{PANAGIO87}
A.Z. Panagiotopoulos, Mol. Phys. {\bf 61},  813  (1987).

\bibitem{Mon1992}
K.K. Mon and K. Binder, J. Chem. Phys. {\bf 96},  6989  (1992).

\bibitem{Bruce97}
A.D. Bruce, Phys. Rev. E {\bf 55},  2315  (1997).

\bibitem{wilding1995a}
N.B. Wilding, Phys. Rev. E {\bf 52},  602  (1995).

\bibitem{frenkelsmit2002}
D. Frenkel and B. Smit, {\em Understanding Molecular Simulation} (Academic, San
  Diego, 2002).

\bibitem{Borgs1992}
C. Borgs and R. Kotecky, Phys. Rev. Lett. {\bf 68},  1734  (1992).

\bibitem{Errington2003}
J.R. Errington, Phys. Rev. E {\bf 67},  012102  (2003).

\bibitem{Binder1981}
K. Binder, Z. Phys. B {\bf 43},  119  (1981).

\bibitem{FISHER98}
M.E. Fisher and S.-Y. Zinn, J. Phys. A {\bf 31},  L629  (1998).

\bibitem{NICOLAIDES88}
D. Nicolaides and A.D. Bruce, J. Phys. A {\bf 21},  233  (1988).

\bibitem{TSYPIN00}
M.~M. Tsypin and H.~W.~J. Bl{\"o}te, Phys. Rev. E {\bf 62},  73  (2000).

\bibitem{GUIDA98}
R. Guida and J. Zinn-Justin, J. Phys. A {\bf 31},  8103  (1998).

\bibitem{bruce1992}
A.D. Bruce and N.B. Wilding, Phys. Rev. Lett. {\bf 68},  193  (1992).

\bibitem{wilding1992}
N.B. Wilding and AD Bruce, J. Phys.: Condens. Matter {\bf 4},  3087  (1992).

\bibitem{KIMFISHER04}
Y.C. Kim and M.E. Fisher, J. Phys. Chem. B. {\bf 108},  6750  (2004).

\bibitem{ferrenberg1989}
A.M. Ferrenberg and R.H. Swendsen, Phys. Rev. Lett. {\bf 63},  1195  (1989).

\bibitem{Sollich_maybe_2010} P. Sollich (unpublished).

\bibitem{KIMFISHER03}
Y.C. Kim and M.E. Fisher, Phys. Rev. E {\bf 68},  041506  (2003).

\end{thebibliography}

\end{document}